\documentclass[largeformat]{interact}

\usepackage{epstopdf}
\usepackage{subfig}

\usepackage[numbers,sort&compress,merge]{natbib}
\bibpunct[, ]{(}{)}{,}{n}{,}{,}

\usepackage{multirow}
\usepackage[utf8]{inputenc}
\usepackage{url}
\usepackage{color,soul}

\usepackage{draftwatermark}
\SetWatermarkText{Draft}
\SetWatermarkScale{5}

\begin{document}

\title{Tracking the evolution of crisis processes and mental health on social media during the COVID-19 pandemic
}

\author{
\name{Antonela Tommasel\textsuperscript{a}\thanks{Corresponding Author. Email: antonela.tommasel@isistan.unicen.edu.ar}, Daniela Godoy\textsuperscript{a} and Juan Manuel Rodriguez\textsuperscript{a}}
\affil{\textsuperscript{a}ISISTAN, CONICET-UNICEN, Argentina
}
}


\maketitle

\textit{Word Count:} $11,501$
\vspace{0.5cm}

\begin{abstract}
The COVID-19 pandemic has affected all aspects of society, not only bringing health hazards, but also posing challenges to public order, governments and mental health. Moreover, it is the first one in history in which people from around the world uses social media to massively express their thoughts and concerns. This study aims at examining the stages of crisis response and recovery as a sociological problem by operationalizing a well-known model of crisis stages in terms of a psycho-linguistic analysis. Based on a large collection of Twitter data spanning from March to August 2020 in Argentina, we present a thematic analysis on the differences in language used in social media posts, and look at indicators that reveal the different stages of a crisis and the country response thereof. The analysis was combined with a study of the temporal prevalence of mental health conversations across the time span. Beyond the Argentinian case-study, the proposed approach and analyses can be applied to any public large-scale data. This approach can provide insights for the design of public health politics oriented to monitor and eventually intervene during the different stages of a crisis, and thus improve the adverse mental health effects on the population.
\end{abstract}



\section{Introduction}

The COVID-19 crisis has affected all aspects of society, not only bringing health hazards, but also posing challenges to public order, governments and mental health. Crisis can serve as both threats and opportunities, as despite the tangible risks to the public, they also draw awareness about the threats, which can be used to steer people towards productive and socially beneficial behaviors~\citep{doi:10.1080/01463373.2015.1012219}. Over the last years, social media has become a part of the daily life of millions of people, as an important medium for exchanging messages in social platforms and for reporting events as they occur. In this sense, the COVID-19 pandemic is the first one in history in which people from around the world have been massively expressing their thoughts and concerns. Hence, there is an unprecedented opportunity to study this pandemic in light of the social media activity it generates, and how the propagation of COVID-related content connects with existing knowledge about crisis processes, mental health and other societal behaviors (e.g., emotions, crime).

Several studies have aimed at identifying, modeling and understanding the varying stages through which crisis arise, evolve and dissipate~\citep{doi:10.1080/01463373.2015.1012219}. These models have been useful for understanding the response efforts, as each stage has its particular needs, thus requiring distinct strategies and resources~\citep{neal1997reconsidering}. Although useful, these models have limitations in practice to explain how particular stages are faced by a social collective or community, during the development of a crisis. A common approach is to study the needs and opportunities of a given community, as manifested by their individuals, government actors and other environmental factors.

In this context, the analysis of the textual exchanges in social media provides rich information of online (and perhaps offline) behaviors of individuals. This analysis can provide insights on: how the perception of the crisis is evolving, how individuals are coping with the crisis, and what their needs are, among others. In addition, social media is changing how people manifest and communicate aspects related to their mental health state. For example, nowadays individuals are more prone to self-identify as suffering from a disorder and to communicate with others sharing similar experiences, which permits to observe the mechanisms underlying mental health conditions during crisis from different perspectives. Similarly, as governments struggle to develop effective messaging strategies to support society, being able to analyze how society perceives and responds to those messages becomes crucial for decision makers.

In this work, we present a study and a supporting approach for characterizing crises like the COVID-19 pandemic based on the usage of language in social media. The approach allows human actors to monitor the evolution of a crisis through its different stages, and eventually plan for interventions, helping to improve the mental health effects of the crisis. To that end, our study aims at examining:
i) the prevalence and evolution of mental health markers, 
ii) the evolution of emotions, and
iii) the stages of crisis response as a sociological problem. 

A key aspect of our approach is the operationalization of a model of crisis stages in terms of lexicons for psycho-linguistic text analysis. Particularly, we performed a thematic analysis on the differences in language use in social media posts with respect to different crisis stages, based on a large collection of Twitter data collected from March to August 2020, containing commonly used hashtags belonged to specific user accounts related to Argentina~\citep{SpanishTweetsCOVID-dataset}. This analysis was combined with a study of the temporal prevalence of mental health conversations (for example, related to depression) across the time span, which shed light on relationship between the crisis stages and the mental health of individuals. Different NLP techniques were used for pre-processing the Twitter data. Furthermore, we developed a heatmap-like visual metaphor for tracking the evolution of the crisis stages as a function of different dimensions of the target model. We believe that this work can contribute to a better understanding of the manifestation of psychological processes related to crisis as they reflect on Spanish-based social media. Thus, it can support the design of public health politics oriented to preserving the mental well-being of individuals during crises. Furthermore, the proposed approach is not tied to the Argentinian case-study, and it can be applied to other large-scale data streams or even incorporate alternative disaster models.

The rest of this paper is organized as follows. Section~\ref{sec:background} presents background concepts of both mental heath and sociological crisis models, as well some related works. Section~\ref{sec:approach} describes the approach for operationalizing the selected psych-social theories and applying it to the collected tweets. Then, Section~\ref{sec:data-analysis} presents the performed analysis. 
Finally, Section~\ref{sec:conclusions} draws the conclusions of this study including its limitations and future lines of work.

\section{Background and related work}\label{sec:background}

Since the beginning of the health crisis due to the COVID-19 pandemic, social media has been a rich source of information for both analyzing the phenomenon and mitigating its effects. 
Datasets of different sizes and characteristics have been released to support the study of the social phenomenon around this pandemic. 
The sampling and collection of social media data, enable researchers (and other actors) to examine different aspects of people's reactions to the pandemic as well as their direct and indirect consequences, as expressed through the use of language in social texts. As an example,~\citet{Li2020} analyzed psychological characteristics in a two-week period before and after the declaration of the COVID-19 outbreak in China on January 20th 2020. Weibo original posts during such period were sampled to explore the impacts of COVID-19 on people's mental health. LIWC~\citep{Pennebaker2001} categories were compared the week before and after the mentioned date, considering categories related to emotions (e.g. positive or negative emotions, anxiety, anger) and concerns (e.g. health, family friends). As expected, the study showed that negative emotions (anxiety, depression, and indignation) and sensitivity to social risks both increased, whereas positive emotions (Oxford happiness) decreased. The focus of concerns also drifted as people were more concerned about health and family, and less about leisure and friends. \citet{Hou2020} also analyzed Weibo posts using LIWC to assess public emotion responses to epidemiological events, government's announcements, and control measures in a period between December 2019 and February 2020. The psycho-linguistic features observed in the study were negative emotions (i.e. anxiety, sad, anger), and risk perception (i.e. drives). Three peaks were reached by all these features during the analyzed period and manifested within 24 hours after the triggering event took place. ~\citet{aiello2020epidemic} empirically tested at scale the model proposed by Strong ~\citep{Strong1990} according to which any new health epidemic resulted into three social epidemics: of fear, moralization, and action. The authors characterized the three social epidemics based on the use of language on social media by means of lexicons and their goal was to embed epidemic psychology in real-time models (e.g., epidemiological and mobility models).

Recently, social media has been also used to understand health outcomes through quantitative techniques that predict the presence of specific mental disorders and symptomatology, such as: depression, suicidality, and anxiety~\citep{Chancellor2020}. Evidence indicates that the rigorous application of even simple natural language processing, computational linguists and psycho-linguistics techniques can yield insights into mental health disorders~\citep{coppersmith-etal-2014-quantifying,10.1145/2464464.2464480,losada2020evaluating}. Some works have explored these techniques in the context of human-made as well as natural disasters. For instance, \citet{Gruebner2017} aimed to identify specific basic emotions from Twitter for the greater New York City area during Hurricane Sandy in 2012. \citet{Lin2014} used geo-coded tweets over an entire month to study how Twitter users from different cities expressed three different emotions (fear, sympathy and solidarity) in reaction to the Boston Marathon bombing in 2013. \citep{dechoudhury2013predicting} created a depression lexicon using a labeled collection of Twitter posts associated to symptoms, among other dimensions. Furthermore, a crisis lexicon called CrisisLex~\citep{olteanu2014crisislex} was created from the sampling of Twitter communications that can lead to greater situational awareness during this kind of crises. These works are mostly oriented to provide real-time tools based on social media to assist during emergency situations and, eventually provide guidelines for first responders.

Disasters and crisis have been described as occurring in phases or stages~\citet{neal1997reconsidering,dewolfe2000training}, which assign order and rationality to the complex reality of disasters and the human responses to them. Phases aim at identifying periods in the unfolding of a crisis, serving to classify the impact or actions that take place to address such impacts~\citep{Kelly1999SimplifyingDD}. Considering a temporal dimension,~\citep{drabek2012human} made a fourfold division of disaster phases: \textit{Preparedness}, \textit{Response}, \textit{Recovery}, and \textit{Mitigation}. The first stage, preparedness, involves actions tending to the elimination or reduction of the effects of a potential disaster. The response phase occurs in the immediate aftermath of a disaster and involves actions in response to challenges caused by disasters (e.g. lack of communications). Then, a phase of recovery takes place in which things starts to return to normal. Mitigation, in turn, refers to sustained actions to reduce or eliminate long-term risks from a disaster occurrence and its effects. Nonetheless, phases should not be considered as discrete events, in which social change is based on a unique episode, but a series of a cycle of events. In this context, the linear division between phases could not represent the reality of every reported or analyzed event~\citep{neal1997reconsidering}, as it might not be easy to find a standard set of measures that identify or quantify how a society evolves through the crisis. In this sense, phases might overlap, and should not include an objective time definition. Instead, as each combination of crisis and society is different, the duration and transition between phases should be adjusted to the “social time”, which accounts for the needs or opportunities of societies~\citep{neal1997reconsidering,dynes1970organized}.

As the stages of crises develop, a range of personal emotions often emerge in response to the situation itself as well as to the social disruption and uncertainty that it causes in people. A crisis not only disrupts the quality of life but also creates a burden of mental health conditions \citep{doi:10.1080/01463373.2015.1012219}, because exposure to a crisis can be a stressor that affects individuals’ expectations about the future, challenging their world views, and triggering emotional reactions~\citep{doi:10.1177/0022167812449190}. 
For example, works as~\citep{Park2012ExploringHO} found initial evidence that people post about their depression on social media, and observed that words about symptoms dominate. The posts often provide details about sleep, eating habits, and other forms of physical ailment, all of which are known to be associated with occurrence of depressive episodes. Likewise, in~\citep{7940253} depression is predicted starting from nine categories associated to possible symptoms, such as: sadness, loss of interest, appetite, sleep, thinking, guilt, tired, movement and suicidal ideation. In this context, social media offers the opportunity of monitoring and understanding the mechanisms underlying mental health conditions during crisis at a massive scale. This kind of monitoring is also the first step to propose actions that can provide “virtual” support to affected individuals. 

From the precedent works, we recognize the potential of social media posts for understanding the different stages of a crisis and mental health-related aspects in a community. One of the challenges here is the bi-directional linkage between a given social theory (or model) and the "reality" as reflected in the texts of the posts, so that the theory becomes actionable for the crisis. On one side, a given theory can provide a structure or framework to reason about the vast flow of posts in social media. On the other hand, the empirical data extracted from the posts can serve to instantiate the different parts of a theory for a given crisis (such as the COVID-19 pandemic), and help up to track the evolution of the phases or stages prescribed by that theory. A systematic analysis of the data can suggest courses of action for managing the crisis, or even signal adjustments to the underlying theory. In this work, we take a step towards this vision by studying a large dataset of COVID-19 tweets, and moreover, by proposing an analysis pipeline that integrates text processing, psycho-linguistic lexicons, and visualization techniques.


\section{Material and Methods}\label{sec:approach}

\begin{figure}
\centering
\includegraphics[width=0.90\columnwidth]{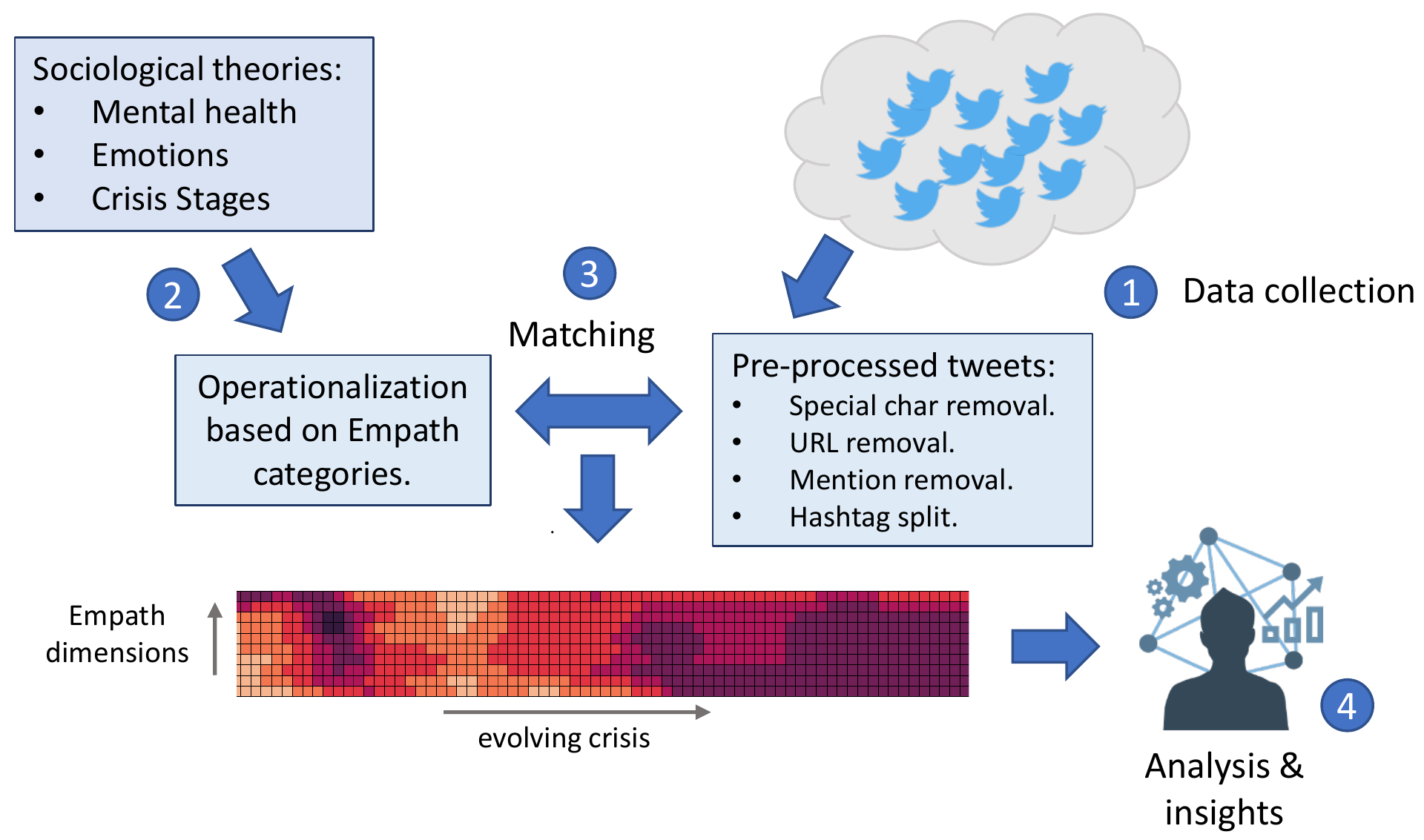}
\caption{Schematic diagram of the proposed approach}\label{fig:approach}
\end{figure}

This section describes the proposed approach for characterizing the crisis phases from language usage on social media for analyzing the prevalence of emotions and mental health discussions. The approach is schematized in Figure~\ref{fig:approach}, and it involves the following steps: 1) data collection and pre-processing, 2) operationalization of sociological theories into lexicons, 3) matching and scoring the pre-processed tweets according to the lexicons, and 4) analysis of the results (time series and heatmaps). 

This approach is guided by three research questions, each one related to whether psycho-linguistic techniques can provide evidence of different aspects of the elements under analysis, i.e. mental health, emotions and crisis.

\begin{itemize}
    \item \textbf{RQ1:} \textit{Mental health}. To what extent social media reflects changes in anxiety, stress and depression markers, as the COVID-19 crisis evolved? Are there changes in the manifestation of the selected disorders?
    \item \textbf{RQ2:} \textit{Emotions}.  How does the prevalence of emotions change as the COVID-19 crisis evolved? Are there any changes in the prevalence of negative or positive emotions? 
    \item \textbf{RQ3:} \textit{Crisis Stages}.  Do crisis stages  manifest on social media? Can a crisis stage of a society be analyzed  
    on the basis of social media behaviors?
    
\end{itemize}

The remainder of this section provides details of the steps of the proposed approach.

\subsection{Data collection}

Analyses are based on the \textit{SpanishTweetsCOVID-19} data collection, which is a large-scale sample of data shared in Twitter during the COVID-19 pandemic in Argentina. We chose Twitter as it is one of the most commonly used social media site, and its role as a public, global and real-time communications provides a glimpse on contemporary society as such~\citep{weller2014twitter}. Twitter additionally enables an easy access to its data, in comparison to data of other social media sites. The collection provides a broad perspective on the dynamics during this health crisis in Spanish-speaking countries, but centered in Argentina. The dataset consists of a collection of Spanish tweets, complemented with geographical information and the possibility of deriving content-based relations between users from the tweet sharing activity. 

\textit{SpanishTweetsCOVID-19} includes more than $150$ million tweets, collected between March 1 and August 30 2020, and it is publicly available in Mendeley\citep{SpanishTweetsCOVID-dataset}\footnote{Available at: \url{https://data.mendeley.com/datasets/nv8k69y59d/1}}. The raw data belonging to the 145 million Twitter posts were retrieved from the Twitter API using the \texttt{Faking it!}\footnote{Available at: \url{https://github.com/knife982000/FakingIt} tool. Since raw data cannot be publicly shared, \texttt{Faking it!} can also be used to rehydrate the data collection.} tool, which internally uses Twitter4J for easily integrating with the Twitter API\footnote{\url{https://developer.twitter.com/en/docs/twitter-api}}.

The collecting process was based on the Twitter Streaming service, which provides real-time access to the shared tweets. The stream was filtered according to the parameters shown in Table~\ref{tab:data_queries}. To be retrieved, tweets had to be identified as written in Spanish, and include any of the selected keywords, or refer to a selected user belonging to the official Argentinian government offices or media. We also considered the retrieval of tweets located inside the Argentina geographical bounding box.

\begin{table}
  \caption{Queries used for collecting the data}
  \label{tab:data_queries}
  \scalebox{0.85}{\begin{tabular}{p{0.3\textwidth}p{0.8\textwidth}}
    \toprule
    Language & es\\
    Keywords (including their hashtag versions) & quedateencasa, covid, covid-19, casa rosada, cuarentena, barbijo, máscara, salud, coronavirus, solidaridad, argentina, caso, muert, infectado, infectada, test, testeo, testeomasivo, tapaboca, médico, enfermera, viruschino, virus, virus chino, jubilado, pandemia, mayorescuidados, tapatelaboca, cuidarteescuidarnos, desarrollo social, cancilleria argentina, argentinaunida  \\
    Users &  LANACION, msalnacion, alferdez, casarosada, infobae, lanacion, clarincom, kicillofok, gcba, felipe\_sola, c5n, mauriciomacri*, cfkargentina*, Chequeado*, DiputadosAR*, HCDiputadosBA*, MDSNacion*, Migraciones\_AR*, MindefArg*, MindeTransporte*, MinSeg*, msnarg*, MinGenerosAR*, MinInteriorAR*, MinTrabajoAR*, msalnacion*, elcancillercom*, filonewsOK*\\
    \bottomrule
      & \footnotesize{* were added to the query in July}\\
  \end{tabular}}
\end{table}

\begin{figure}[t]
\centering
\includegraphics[width=0.90\columnwidth]{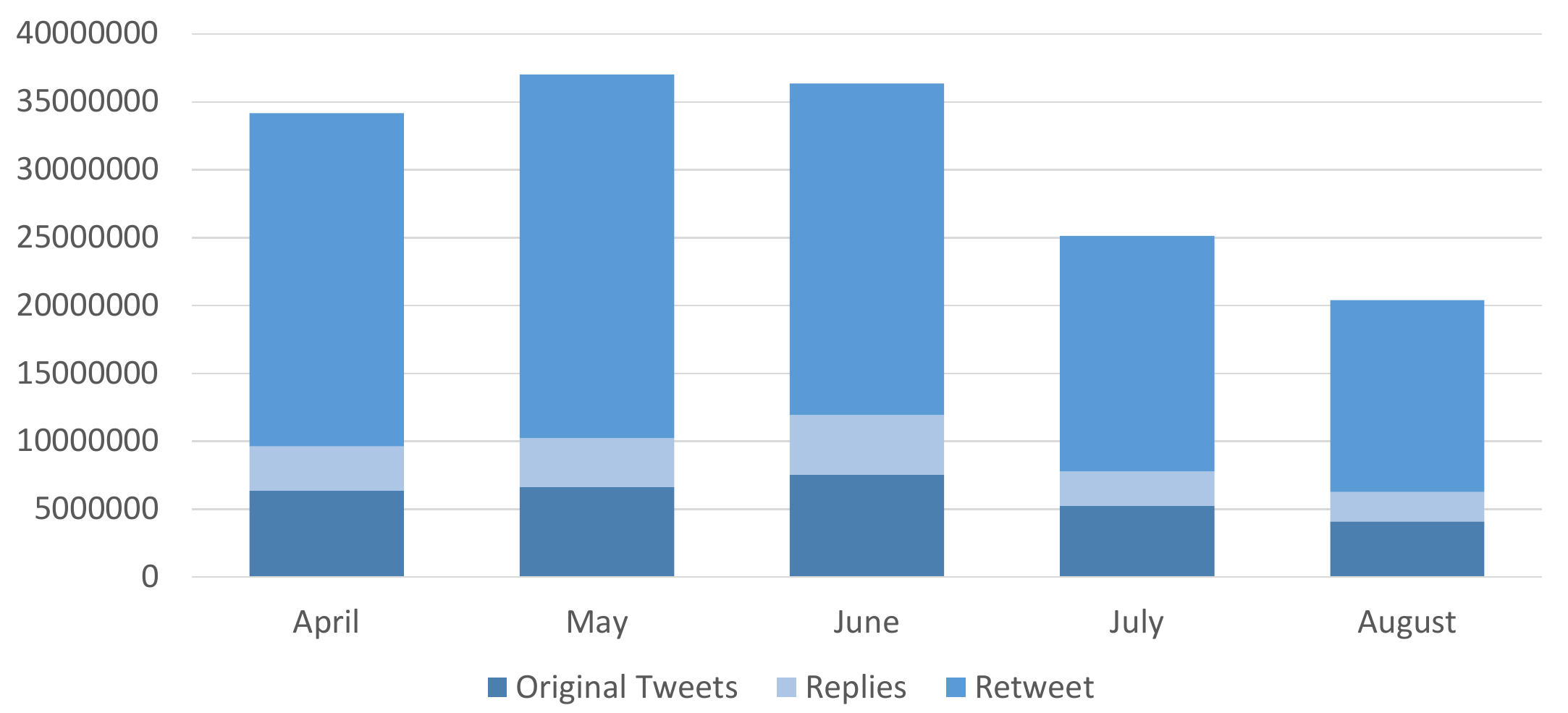}
\caption{Distribution of tweets per month and type}\label{fig:data_collection_monthly}
\end{figure}

Figure~\ref{fig:data_collection_monthly} shows the monthly distribution of the tweets collected using the queries in Table~\ref{tab:data_queries} and their type (original, retweets and replies). The number of tweets in March represents less than $1\%$ of the total collected tweets, thus it is not included in the figure. As it can be observed, the original tweets were only the $19\%$ of the dataset, whilst retweets accounted for the $70\%$. The remaining tweets correspond to replies. Interestingly, the number of collected tweets reached its peak in June. For the purpose of our analyses, only original tweets and replies were considered, as they are the ones in which writers provide sufficient content so as to evaluate aspects related to the crisis evolution and mental health. 
Table~\ref{tab:data_stats} summarizes the statistics of the collected dataset, excluding retweets. Hashtags are only present in the $19\%$ of the tweets. An inspection of the data showed that neither official government accounts, politics nor media included hashtags in the majority of their tweets. The median number of shared tweets per user was $2$, whilst the $40\%$ of the accounts tweeted over the median.

\begin{table}[t]
  \caption{Statistics of the collected dataset}\label{tab:data_stats}
\begin{tabular}{cccc}
\toprule 

\multirow{5}{*}{{\small{}Tweets}} & \multicolumn{3}{p{0.75\columnwidth}}{{\small{}Start date: 1 March 2020 -- End date: 31 August 2020}}\tabularnewline

\cmidrule{2-4} \cmidrule{3-4} \cmidrule{4-4} 
 & {\small{}Original: 29,712,138} & {\small{}Retweets: 106,361,088} & {\small{}Replies: 15,815,975 }\tabularnewline
 & {\small{}(with quotes: 8,947,933)} &  & {\small{}(with quotes: 414,984)}\tabularnewline
\cmidrule{2-4} \cmidrule{3-4} \cmidrule{4-4} 

& {\small{}With media: 7,578,945 } & \multicolumn{2}{l}{{\small{}With mentions: 19,494,778}}\tabularnewline
\cmidrule{2-4} \cmidrule{3-4} \cmidrule{4-4} 
 & {\small{}With URL: 9,917,012} & \multicolumn{2}{l}{{\small{}With places: 102,241 }}\tabularnewline
\cmidrule{2-4} \cmidrule{3-4} \cmidrule{4-4} 
 & {\small{}With hashtags: 8,621,975} & \multicolumn{2}{p{0.5\columnwidth}}{{\small{}Avg. number of hashtags per tweet: 46}{\small\par}
{\small{}Avg. frequency of hashtags: 13}}\tabularnewline

\midrule 
\midrule 
\multirow{2}{*}{{\small{}Users}} & {\small{}Total: 6,977,032} &  & \tabularnewline
\cmidrule{2-4} \cmidrule{3-4} \cmidrule{4-4} 
 & {\small{}Min tweets per user: 1} & {\small{}Avg. tweets per user: 4.5 (sd: 8.4)} & {\small{}Max tweets per user: 167}\tabularnewline

\bottomrule
\end{tabular}
\end{table}

The collected tweets were pre-processed to remove special characters, URL and mentions. Hashtags were kept without the numeral symbol and were split into their constituent words. After an inspection of randomly selected tweets, spelling corrections were not applied as most of the tweets were correctly written, and we did not detect an abuse of abbreviations. As our approach relies on word matching, and the original Empath lexicon was not lemmatized, we did not apply neither lemmatization nor stemming.

\subsection{Operationalization of psychological theories}

The language used in social media for expressing opinions, personal situations and communicating with others provides signals about a person state of mind and situation. In this sense, lexicons are a rich tool for analyzing language in social texts across a broad range of categories, including emotions, concerns and health-related issues. Existing language lexicons
have been widely used in psychometric studies as well as sentiment analysis. Some examples are the Linguistic Inquiry and Word Count (LIWC)~\citep{Pennebaker2001}, Emolex~\citep{MohammadT13}, the Prosocial Behavior Lexicon~\cite{Frimer2014} and LEW list~\citep{Francisco2013}, among others. When no training data is available, the availability of domain-specific lexicons plays a fundamental role on the automated analysis of texts. Thus, accurate lexicons can be of valuable guidance to understand the person behind the text.

In this study, the psycho-linguistic analysis of social texts was based on the multiple categories provided by two specific lexicons:

\begin{itemize}
\item \textit{Empath}~\cite{empath1,empath2} covers a broad, human-validated set of 200 emotional and topical categories drew from common concepts in the ConceptNet~\cite{Liu2004} knowledge base and Parrott's hierarchy of emotions~\cite{Parrott2001}. Categories have a number of seed terms representing the concept, which can be further expanded to obtained similar categorical terms. Empath categories have been shown to be highly correlated with similar categories in LIWC and EmoLex.
\end{itemize}
\noindent Considering the positive experiences reported by~\citet{perczek2000coping}, who analyzed the reliability of Spanish translation from English for psychometric scales and lexicons, as the lexicon is only available in English, we automatically translated the words associated to the categories using the IBM Watson Language Translator\footnote{\url{https://www.ibm.com/watson/services/language-translator/}}. Then, each of the authors independently checked the translation for inconsistencies. In addition, considering the gendered nature of Spanish, in case a word in a category is associated to a specific gender, we added the words corresponding to the other gender, and also a gender-neutral version (if existed). 

\begin{itemize}
\item \textit{SentiSense Affective Lexicon}~\cite{de-albornoz-etal-2012-sentisense,Carrillo-de-AlbornozCA12}  consists of synsets from WordNet\footnote{\url{https://wordnet.princeton.edu/}} labeled with an emotional category. Particularly, this lexicon consists of 2,190 synsets labeled with 14 emotional categories derived from the ones proposed by~\citet{Arnold1960},~\citet{Plutchik1980} and~\citet{Parrott2001}.
\end{itemize}
\noindent  Since SentiSense Affective Lexicon is integrated with the WordNet Spanish version, it can be directly applied to the analysis of our tweets. These categories are the ones used for the analysis of emotions.
\vspace{3mm}

The operationalization of the crisis stage and mental health theories was accomplished in three steps, namely: i) adaptation and expansion of lexicons based on a thematic encoding and categorization; ii) contextualization of the expanded lexicons based on semantic similarity; and iii) association between the expanded lexicons and the Empath categories that will later be used in the analysis of the pre-processed tweets. The contextualization of the expanded lexicons was based on \textit{FastText}\footnote{\url{https://fasttext.cc/}} embeddings. The goal here was to capture the context in which the words in the lexicon appear, in order to gather a more accurate picture of how individuals express and manifest aspects related to crisis evolution and mental health. Then, for each word in our lexicons, we selected the top-10 most similar terms in FastText, based on the traditional Wikipedia model. Finally, we matched each of the expanded terms to the Empath categories to automatically retrieve the top-10 most prevalent categories for each disorder, i.e. the categories with the highest number of shared words with the lexicons. We refer to the selected categories as \textit{markers} for the associated lexicon.
 
The original lexicons, expansions and the translation of Empath categories can be found it the corresponding companion repository\footnote{Due to the requirement of anonymous submission the figures are momentarily located in the following Drive folder: \url{https://tinyurl.com/yymn3e72}}. The following subsections describe the operationalizations of the mental health and crisis stages.
 
\subsubsection*{Mental Health}

In public mental health terms, the main psychological impact of the COVID outbreak have been elevated rates of stress and anxiety~\citep{asmundson2020health}. However, as a consequence of public health policies, such as social distancing and the uncertainty generated by social and economic situations, a rise on depression, levels of loneliness, addictions and suicidal behaviors can be expected. Thus, in this study we focused on anxiety, stress and depression as the three main concerns regarding mental health in the population. 

As previously mentioned, several works~\citep{Park2012ExploringHO, 7940253, dechoudhury2013predicting} have relied on lexicons to automatically analyze the level of depression in texts from social networks. Based on such lexicons, each of the authors manually hand-coded the characterizations, symptoms and manifestations of each selected disorder as defined by the National Institute of Mental Health\footnote{\url{https://www.nimh.nih.gov/index.shtml}} and the Anxiety and Depression Association of America\footnote{\url{https://adaa.org/}}. This coding generated independent lists of keywords that were combined using a voting strategy. Then, the resulting lexicons were augmented using FastText to not only included descriptions of the manifestations of the selected mental health disorders, but also, additional words that are commonly used or provide context to the manifestations, as in~\citep{losada2020evaluating,lin2016does}. Finally, we matched each of the expanded lexicons to the Empath categories.  Table~\ref{tab:mapping-mental-health} shows a summary of the words associated to each disorder, and the corresponding Empath categories.


\begin{table}[t]
\caption{Mapping of mental health issues to \textit{Empath} categories\label{tab:mapping-mental-health}}

\begin{tabular}{p{0.1\columnwidth}p{0.6\columnwidth}p{0.2\columnwidth}}
\toprule 
{\footnotesize{}Mental disorder}{\footnotesize\par} & {\footnotesize{}Associated keywords}{\footnotesize\par} & {\footnotesize{}Associated categories}{\footnotesize\par}
\midrule
{\footnotesize{}Anxiety} & {\footnotesize{}anxiety, medication, depression, meds, panic attack, falling, heart, fear, apprehension, nervousness, restlessness, suffering, uncertainty, unease, worry, tension, wound-up, frustrated, irritability, sleep problems, pounding heartbeat, sweating, trembling, shaking, nightmare, frightening thoughts, angry outburst, trauma, loss of interest, eating habits, demoralize}{\footnotesize\par} & {\footnotesize{}sadness, nervousness, fear, suffering, horror, disappointment, health, confusion, shame, anger}{\footnotesize\par}
\midrule 
{\footnotesize{}Depression} & {\footnotesize{}depression, boring, sadness, painful, unhappy, suicide, dissatisfaction, confused, unsatisfactory, cry, die, hopeless, indecisive, impatience, reluctant, fatigue, palpitation, useless, underestimated, disappointed, withdrawal, insomnia, drowsiness, dizziness, nausea, seizures, antidepressant,medication, apathetic, demanding, guilty, shame, remorse, demands, loss of satisfaction, complaining, tachycardia, fatigue, emptiness, restless, pain,sedative, unhappiness, detach}{\footnotesize\par} & {\footnotesize{}sadness, suffering, shame, neglect, emotional, disgust, torment, nervousness, disappointment, pain}{\footnotesize\par}
\midrule 
{\footnotesize{}Stress} & {\footnotesize{}stress, agony, anxiety, burden, fear, intensity, nervousness, tension, worry, affliction, apprehensiveness, distension, fearfulness, impatience, mistrust, nervous, interruptions, expectations, agonized, alienation, alone, anger, angry, anguish, antidepressant, disinterest, distracted, done with life, doomed, down, downhearted, drained, drugs, hopeless, fatigue, fear, pessimistic, terrified}{\footnotesize\par} & {\footnotesize{}sadness, nervousness, anger, suffering, fear, shame, torment, neglect, disgust, health}{\footnotesize\par}
\bottomrule
\end{tabular}

\end{table}

\subsubsection*{Crisis stages}
Several authors have studied the sociological responses to natural disaster events~\citep{drabek2012human,Richardson2005ThePO}, based on surveys, interviews and narratives. Nowadays, social media presents new ways to explore communication during a crisis
~\citep{10.1371/journal.pone.0210484}. Following the crisis stages defined by \citet{neal1997reconsidering}, we attempted to characterize them based on the communicative ways in which the social collectives organize themselves~\citep{Richardson2005ThePO}.

Departing from the crisis stages defined by \citet{drabek2012human}, \citet{Richardson2005ThePO} and \citet{dewolfe2000training}, as well as the crisis lexicon proposed by~\citet{olteanu2014crisislex} each of the authors manually hand-coded each of the described stages. We obtained keywords related to the four traditional crisis stages, namely: preparedness (including the aspects related to planning and warning), response (including the aspects related to impact, the heroic and disillusionment sub stages), recovery and mitigation. From the crisis lexicon we only kept those words related to actions, sentiments, and emotions, removing all words that were related to a particular type of crisis (e.g. tornado, flood, explosion, storms, among others). The selected keywords include not only aspects related to the functional aspects for each of the stages by the different actors (e.g. individuals, society, government), but also aspects related to the perceptions and concerns during a crisis (e.g. the false sense of security that is commonly felt during the preparedness phase), and aspects related to mental health. 
As previously mentioned, the resulting lexicons were augmented using FastText and matched to the Empath categories to retrieve the 10 most prevalent categories for each disorder.

Table~\ref{tab:mapping-crisis} shows a summary of the words associated to each disorder, and the corresponding Empath markers. As it can be observed, some markers are shared across stages, meaning that the concepts or keywords that describe each stage might be relevant to the others. 
This implies  that stages are not linearly divided, as \citet{neal1997reconsidering, Kelly1999SimplifyingDD} stated. 

\begin{table}
\caption{Mapping of crisis stages to \textit{Empath} categories\label{tab:mapping-crisis}}
\centering
\scalebox{0.98}{
\begin{tabular}{p{0.1\columnwidth}p{0.6\columnwidth}p{0.2\columnwidth}}
\toprule 
{\footnotesize{}Crisis stage}{\footnotesize\par} & {\footnotesize{}Associated keywords}{\footnotesize\par} & {\footnotesize{}Associated categories}{\footnotesize\par}
\midrule
{\footnotesize{}Preparedness}{\footnotesize\par} & {\footnotesize{}government, responsibility, society, public belief,
we, faith in authority, lack of concern, family, curiosity, safety,
acceptance, anticipation, resources, response, anticipation, hoard,
preparedness, think, anxiety, worry, futility, helplessness, protective,
control, overwhelmed, warning, impact, unreal, sense of security,
information, track, control, reality, plan, devise, strategy, idea,
liability, importance, duty, care, charge, restrain, power, prospect,
contemplation, expectancy, premonition, promise, presentiment, preoccupation,
sensation, change, defensive, avoidance, realistic, vigilance, disbelief,
normality bias, interpretations, naturalize the threat, denial, danger,
information, perception, diversity, confirmation efforts, misinterpretation,
refusal, warning, threat, adaptation, panic, consequences, credibility,
vulnerable, vulnerability, unsafe, fearful, no control, protection,
guilt, self blame, advice, alarm, alert, caution, guidance, indication,
notification, avoidance, panic, ready, alertness, preparedness, forewarning,
expectation, provision, awareness, prevision, foreseeing, consequence,
advice, recommendation, suggestion, forecast, attention}{\footnotesize\par} & {\footnotesize{}fear, sadness, nervousness, horror, neglect, aggression,
anticipation, disappointment, communication, trust}{\footnotesize\par}
\midrule
{\footnotesize{}Response} & {\footnotesize{}fear, hoarding, credibility, active, anxiety, entrapment,
powerlessness, impotence, isolation, dependency, adaptive, responsible,
emotional disturbance, dazed, stunned, apathetic, passive, immediate,
extreme, gratitude, help, concern for family and community, susceptibility,
irritability, anger, angry, sadness, withdrawal, obsessive preoccupation,
safety, loss of interest, depression, resistance to authority, feeling,
inadequacy, panic, afraid, help, medication, depression, mental, mind,
way, life, time, side effects, meds, music, nervousness, restlessness,
suffering, uncertainty, unease, worry, sleep problems, sweating, trembling,
shaking, shortness of breath smothering, chocking, being out of control,
frightening thoughts, trigger, easily startled, angry outburst, anger,
trauma, stress, loss of interest, angst, sadness, concern, disinterest
in life, distress, fearful, unknown, hazard, concerned, demoralize,
shock, disconcerting, safety, rapport, government, volunteer, aid,
attend, charge, repair, palliate, reform, limits of assistance, mental
health, exhausted, exhaustion, stress, discouragement, fatigue, problems,
frustration, antidepressant, medication, shame, remorse, lonely, lonesome,
inept, inhuman, insensate, irritable, intense, intolerable, miserable,
grief, rage, purposelessness, upset, disorientated, i, me}{\footnotesize\par} & {\footnotesize{}{sadness, suffering, nervousness, shame, neglect, disgust, fear, anger, health, disappointment}}{\footnotesize\par}
\midrule 
{\footnotesize{}Recovery} & {\footnotesize{}victim, perception, recovery, disengagement, recurrence,
fear, anxious, legal implications, financial state, depression, shocked,
distress, demoralization, overacting, loss of appetite, difficulty
in sleeping, shaky, trouble, irritability, sweating, anger, headache,
stomach trouble, diarrhea, constipation, alcohol, tobacco, smoking,
drugs, pills, tranquilizers, insomnia, restless, sleep, dependency,
clinging, despair, preoccupation, sensitivity, death, hostility, apathy,
home, enclosed, people, street, irritation, living, agitation, tears,
sorrow, worry, recover, migraine, burden, sarcasm, jokes, guilt, tension,
inept, government, politics, conflicts, restrictions, breakup, attention
to present, leveling of society, i, we, community identification,
society, community, looting, robbery, death, grief, nightmares, need
to talk, crime, talk, optimism, religion, guilt, phobias, planning,
explanation, increase action, local governments, power, law, economics,
normal, restoration, future, improving, renewal, resurgence, revival,
blame, who, blame}{\footnotesize\par} & {\footnotesize{}sadness, suffering, disgust, anger, nervousness, irritability,
disappointment, fear, neglect, rage}{\footnotesize\par}
\midrule 
{\footnotesize{}Mitigation} & {\footnotesize{}awareness, salience, underestimate, stability, experience}{\footnotesize\par} & {\footnotesize{}love, anticipation}{\footnotesize\par}

\bottomrule
\end{tabular}
}
\end{table}

\subsection{Matching the tweets and categories}

In the context of our research questions, we analyzed the prevalence of the Empath categories (or markers) associated to each lexicon, by matching the categories to the text in each tweet. A tweet was considered to match a marker if at least one word belonged to such marker. This matching does not consider the number of matching words in tweets, because multiple occurrences might be rare due to the restrictions on tweets length, which might not adequately reflect the actual intensity of the category. To summarize the prevalence of a category on a particular day, we computed the percentage of tweets on such day that matched with such category. Once we computed the matching for a marker over the full-time span, we obtained its time series distribution.

In average, in the collected dataset  $189.260$ tweets were shared per day, with a maximum monthly average of $396.228$ in June, and a minimum average of $800$ in March. To reduce the impact of day-to-day variations and weekly periodicity in the obtained time series, and thus better expose the characteristics of the time series, we applied a smoothing considering a window of one week. This smoothing causes the time series to respond more slowly to recent changes, which in turn favors the observation of more consistent behaviors over longer periods of time (in opposition of instantaneous shifts).

Given the smoothed time series, we proceeded to identify phases characterized by the different subsets of Empath markers associated to each of the lexicons. To do so, we searched the time series for break points or peaks, which are defined as points in time in which the values of the involved categories varied altogether. These variations were presumably caused by COVID related events. Over the identified peaks, we only kept those whose prominence values was higher that the mean plus one standard deviation~\citep{palshikar2009simple}. Due to the smoothing, the events leading to the peaks should not be searched on the exact day the peak appears, but also on the events of the days leading to the peak. Peaks were not computed over the smoothed time series, but over the smoothed gradient of the time series distribution. Gradients allowed us to measure magnitude of change (either an increment or decrement) of the time series. 

\section{Data Analysis}\label{sec:data-analysis}

This section presents the psycho-linguistic analysis of: i) the prevalence and evolution of mental health markers, ii) the evolution of emotions, and iii) the stages of crisis response as a sociological problem.

\subsection{Mental Health}

The first research question is about the extent to which tweets contain references to mental health problems, how such references evolve over time with the COVID-19 pandemic, and whether perceptible changes could be observed in the prevalence of the categories associated to the mental health problems. In particularly, we analyzed references to \texttt{anxiety}, \texttt{depression} and \texttt{stress}, which are three of the most commonly analyzed mental illnesses or disorders. 

\begin{figure}
 \centering
   \subfloat[Anxiety\label{fig:mental-health-anxiety}]{\includegraphics[width=0.98\columnwidth]{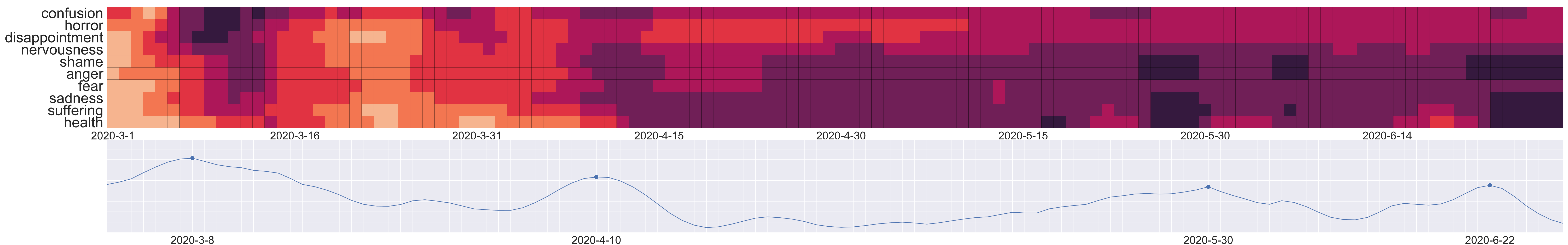}}\\
   \subfloat[Depression\label{fig:mental-health-depression}]{
     \includegraphics[width=0.98\columnwidth]{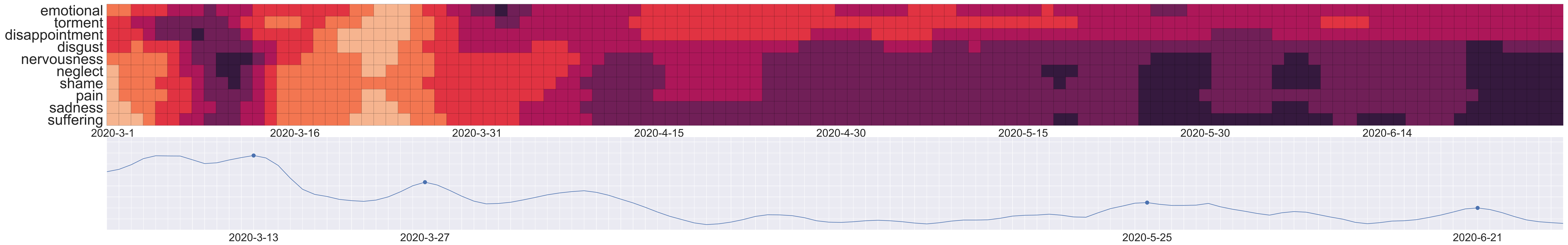}}\\
   \subfloat[Stress\label{fig:mental-health-stress}]{\includegraphics[width=0.98\columnwidth]{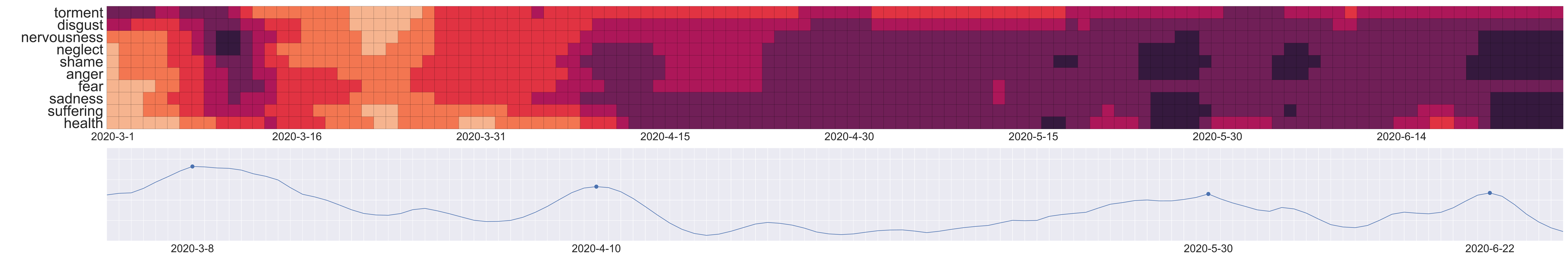}  }
   
 \caption{Prevalence of the Empath categories for the three analyzed mental health disorders}\label{fig:mental-health}
\end{figure}

Figure~\ref{fig:mental-health} presents the temporal distribution of the Empath categories (i.e. markers) for the three disorders during the span March-June, as well as the peaks detected using such  markers. The darker the area the higher the prevalence of the associated markers. The analysis does not include the span July-August due to the different orders of magnitude across the prevalence of markers\footnote{The figures including a complete time span can be found in the companion repository.}. As Argentina moved into 100 days of lock-down (late-June), the markers showed an increment in their prevalence, which hindered the analyses and eclipsed the changes in the preceding months. Table~\ref{tab:peaks-events} shows the correspondence between the discovered peaks, and events in Argentina related to the COVID-19 pandemic and the accompanying political and economic situation.

\begin{table}

\caption{Events associated to peaks observed for the mental health analysis\label{tab:peaks-events}}
\scalebox{0.90}{

\begin{tabular}{p{0.1\columnwidth}p{0.9\columnwidth}}
\toprule 
{\footnotesize{}Peaks}{\footnotesize\par} & {\footnotesize{}Associated events}{\footnotesize\par}
\midrule
{\footnotesize{}March 8th} & \textbf{\footnotesize{}March~3rd}{\footnotesize{}. The first COVID-19 case is confirmed in a man who had arrived from Italy. }{\footnotesize\par}

\textbf{\footnotesize{}March~4th.}{\footnotesize{} The Health Minister
announces that the government was not planning on closing schools,
suspending shows nor flights coming from affected countries. }{\footnotesize\par}

\textbf{\footnotesize{}March~5th.}{\footnotesize{} The Health Minister
declares that people were over-reacting. The flag carrier (Aerolíneas
Argentinas) announced the cancelation of flights connecting Argentina
and Italy. }{\footnotesize\par}

\textbf{\footnotesize{}March~7th.}{\footnotesize{} The first death
is confirmed. A 64-year-old man who had travelled to Paris. The COVID
case was diagnosed post-mortem. }{\footnotesize\par}

\textbf{\footnotesize{}March~8th.}{\footnotesize{} Twelve confirmed
cases.}{\footnotesize\par}
\midrule 
{\footnotesize{}March 13th} & \textbf{\footnotesize{}March 9th.}{\footnotesize{} Agricultural producers
go on strike due to new taxation policies. The president declares
that the strike is violent. The president declares that COVID-19 is
posing a serious situation, but people should not over-react. The
president allows the start of debt restructuring negotiations. }{\footnotesize\par}

\textbf{\footnotesize{}March 11th.}{\footnotesize{} The president
establishes a lock-down only for those returning from an affected
country or region, and affirms that those who break the lock-down
will be prosecuted. }{\footnotesize\par}

\textbf{\footnotesize{}March 12th}{\footnotesize{}. The president
announces the establishment of repatriation mechanisms. }{\footnotesize\par}

\textbf{\footnotesize{}March 13th.}{\footnotesize{} The second death
is confirmed.}{\footnotesize\par}
\midrule 
{\footnotesize{}March 27th} & \textbf{\footnotesize{}March 21st.}{\footnotesize{} According to the
president, "the worst is yet to come". }{\footnotesize\par}

\textbf{\footnotesize{}March 22nd.}{\footnotesize{} Governors asks
for the army to control the lock-down. }{\footnotesize\par}

\textbf{\footnotesize{}March 24th.}{\footnotesize{} Riots in several
prisons. More agreements regarding the operations of the legislative
branch. }{\footnotesize\par}

\textbf{\footnotesize{}March 25th.}{\footnotesize{} The president
threats companies that do not respect the maximum price limits. }{\footnotesize\par}

\textbf{\footnotesize{}March 26th.}{\footnotesize{} The aerial space
is closed. All borders, airports and ports are closed. Not even the
Argentinian people can enter the country.}{\footnotesize\par}

\textbf{\footnotesize{}March 27th.}{\footnotesize{} The president
declares that he is not in a rush to reopen schools. }{\footnotesize\par}
\midrule 
{\footnotesize{}April 10th} & \textbf{\footnotesize{}April 4th.}{\footnotesize{} Mercosur approves
an emergency fund and 45.000 diagnostic tests will arrive. }{\footnotesize\par}

\textbf{\footnotesize{}April 5th.}{\footnotesize{} A TV special ("Unidos
por Argentina", "United for Argentina")
is held to collect donations through the Red Cross. More than 1500
confirmed cases. }{\footnotesize\par}

\textbf{\footnotesize{}April 6th.}{\footnotesize{} Controversy over
purchases made by the government with surcharges. }{\footnotesize\par}

\textbf{\footnotesize{}April 8th.}{\footnotesize{} More than 800 prisoners
are benefited with house arrest for fear of contagions in Provincia
de Buenos Aires. The first national diagnostic tests are being developed. }{\footnotesize\par}

\textbf{\footnotesize{}April 10th.}{\footnotesize{} Second lock-down
extension until April 26rd. Argentina is in the top-10 countries with
the most cases in Latin America.}{\footnotesize\par}
\midrule 
{\footnotesize{}May 25th} & \textbf{\footnotesize{}May 18th.}{\footnotesize{} Presentation of
a new diagnostic kit locally produced. }{\footnotesize\par}

\textbf{\footnotesize{}May 19th.}{\footnotesize{} New activities are
authorized in Buenos Aires. Definitions regarding the authorizations
of new activities. }{\footnotesize\par}

\textbf{\footnotesize{}May 21st}{\footnotesize{}. Negotiations to
avoid debt default. Record of contations in Buenos Aires and CABA. }{\footnotesize\par}

\textbf{\footnotesize{}May 23rd.}{\footnotesize{} Limits to the public
transportation system in CABA. Argentinian GDP is expected to slump
up to a 20\%. Lock-down extended two extra weeks. }{\footnotesize\par}

\textbf{\footnotesize{}May 24th.}{\footnotesize{} The government started
to provide economic assistance to companies. More than 12,000 confirmed
cases.}{\footnotesize\par}

\textbf{\footnotesize{}May 25th.}{\footnotesize{} The president declares
that all economic problems are due to the pandemic and not the imposed
lock-down. Commemoration of first independent government in Buenos
Aires.}{\footnotesize\par}
\midrule
{\footnotesize{}May 30th} & \textbf{\footnotesize{}May 24th.}{\footnotesize{} The government started
to provide economic assistance to companies. More than 12,000 confirmed
cases. }{\footnotesize\par}

\textbf{\footnotesize{}May 25th.}{\footnotesize{} The president declares
that all economic problems are due to the pandemic and not the imposed
lock-down. Commemoration of first independent government in Buenos Aires. }{\footnotesize\par}

\textbf{\footnotesize{}May 29th.}{\footnotesize{} A group of intellectuals
declares that "democracy is in danger",
which causes counter declarations from the government.}{\footnotesize\par}

\textbf{\footnotesize{}May 30th.}{\footnotesize{} Argentina continues
with debt restructuring negotiations to avoid default. Health workers
protest against work insecurity and low salaries.}{\footnotesize\par}
\midrule 
{\footnotesize{}June 21st\footnotesize\par June 22nd} & \textbf{\footnotesize{}June 15th.}{\footnotesize{} The governor of
Provincia de Buenos Aires wants to impose an even harsh lock-down,
as according to estimations, the health system will collapse in 35
days. The economy ministry is confident on reaching an agreement for
the debt restructuration. }{\footnotesize\par}

\textbf{\footnotesize{}June 16th.}{\footnotesize{} The government
suggests that the lock-down might be in place for at least three more
months. }{\footnotesize\par}

\textbf{\footnotesize{}June 17th.}{\footnotesize{} Restrictions for
running according to the number of the national ID. New circulation
permit in place. The president blames the runners for the new surge
of contagions. }{\footnotesize\par}

\textbf{\footnotesize{}June 20th.}{\footnotesize{} Protestors against
the expropriation of a company. }{\footnotesize\par}

\textbf{\footnotesize{}June 22nd.}{\footnotesize{} A plan for restricting
the lock-down is announced. Almost 50k confirmed cases. }{\footnotesize\par}
\bottomrule
\end{tabular}
}

\end{table}

As regards \texttt{anxiety} (Figure~\ref{fig:mental-health-anxiety}), the time series shows the apparition of darker areas in the days following the confirmation of the first COVID case and previous to the suspension of activities and the official declaration of lock-down (between March 7th and March 14th), particularly affecting the markers of \textit{confusion}, \textit{horror}, \textit{disappointment} and \textit{nervousness}. On the other hand, the less affected categories are \textit{suffering} and \textit{health}. This correlates to the apparition of the first peak around March 8th. The burst in \textit{anger} in early March could have been caused by a political event related to the COVID by which the executive branch of government was assigned “special faculties” to discretionally dictate acts of law and make budget allocations. This was also accompanied by declarations of the Health Minister underestimating the virus. These observations match the ones made by~\citet{aiello2020epidemic} after the detection of the first contagion in the US, when there was a peak of \textit{anger}, \textit{fear} and \textit{anxiety}, which decreased after the declaration of lock-down.

After the announcement and start of the first lock-down (March 20th – March 21st), it can be observed that the least prevalent marker was \textit{disappointment}, and a mild decrease in the intensity of  \textit{nervousness}, \textit{sadness}, \textit{suffering} and \textit{health} related tweets. An analysis of the tweets corresponding to such days, shows the faith that people had on the government decisions, the believe on lock-down as something positive, and how grateful they were for how the government was taking care of the population. 
These observations agree with the results of a survey~\footnote{\url{https://cordoba.conicet.gov.ar/wp-content/uploads/sites/25/2020/04/Informe-final-6-de-abril-recortado.pdf.pdf}} collected in Argentina at the end of March, in which people reported an increased sense of security against COVID and a higher sense of awareness regarding the prevention actions, when compared to the start of lock-down. 
Then, as the lock-down continued, we can observe an increment in the conversations related to \textit{health} and \textit{suffering}, followed by \textit{confusion}, \textit{fear} and \textit{anger}. Reaching the 100 days of lock-down (June 27th), the prevalence of \textit{suffering}, \textit{health} and \textit{anger} reached their highest values for a continuous week.

In a context with high \texttt{anxiety} markers, the increment in the prevalence of the \textit{health} category could be related to the phenomenon known as “health anxiety”~\citep{asmundson2020health}. This phenomenon arises from the misinterpretation of perceived body sensations and changes in combination with the consumption of inaccurate or exaggerated information from the media~\citep{asmundson2020coronaphobia,asmundson2020health}. These observations are in accordance with~\citet{asmundson2020health}, who reported an increment in health anxiety, particularly in areas where the number of people affected by COVID-19 were continuously increasing. At the individual level, health anxiety can manifest as maladaptive behaviors (e.g., the avoidance of health care even with genuine symptoms~\footnote{In this regard, the official government communications asked people to stay at home and avoid attending to hospitals, which resulted in the population avoiding hospitals even when showing symptoms, and hospitals suspending services.}, or the hoarding of particular sanity items\footnote{To avoid this, in mid-April, the government fixed the prices of hand sanitizer.}). Then, at a society level, it can lead to mistrust of public authorities, which in turn, can influence the success or failure of the public health strategies put in place. In this context, it is critical for decision makers to understand how health anxiety can influence the responses of individuals and society to the health recommendations~\citep{rajkumar2020covid}.  

As Table~\ref{tab:peaks-events} shows, the apparition of \texttt{anxiety} peaks could be explained by a sequence of events. Most of them are related to the situation of the capital city of Argentina (called Ciudad Autónoma de Buenos Aires, CABA) and the biggest state (called Buenos Aires), which defined the lock-down path for the rest of the country. In the mapping, there also appear events related to debt negotiation (which altered the rate exchange of the ARS and thus good prices), and to citizen manifestations in reaction to decisions of the executive branch of government. When individually analyzing the markers, it can be observed that their peaks, although appear on similar time spans, they do not share the same prominences. Instead, the \textit{health} marker presents the most prominent peaks, followed by \textit{fear}.

When observing \texttt{depression} (Figure~\ref{fig:mental-health-depression}), the tendencies are similar to those of \texttt{anxiety} due to the shared markers. Nonetheless, the distributions show a higher prevalence of most categories since early-April, when it was announced that the lock-down was extended for 15 extra days and that the allowed activities would be limited. 
The topics discovered 
around those days express concerns regarding both health and the economic situation. The hashtag “\#quedatenecasa” (stay at home) still appears on the discovered topics. The highest prevalence is observed for \textit{sadness} and \textit{suffering} across most of the time span. 
When reaching the 100 days under lock-down in late-June, it can be observed a high prevalence of the $60\%$ of categories. On the contrary, \textit{emotional}, \textit{torment} and \textit{disappointment} did not show a high prevalence across the time spam. The first lock-down and its extensions were the most restrictive in terms of the activities that were allowed at a country level. In this sense, the population started to manifest their discouragement regarding the adopted measures.  

The peaks for \texttt{depression} are close to those of \texttt{anxiety}, with the addition of a new one on March 27th. As Table~\ref{tab:peaks-events} shows, that date marks the first week of lock-down and a televised announcement of the president in which he declared the first lock-down extension for two additional weeks. When individually analyzing the markers, both \textit{suffering} and \textit{sadness} showed the most prominent peaks. 

Finally, as regards \texttt{stress} (Figure~\ref{fig:mental-health-stress}), it can be observed the highest prevalence of most of the categories for the longest time span. The categories associated to this disorder match subsets of the categories associated to both anxiety and depression, with the addition of the stress category. As for depression, the detected peaks match those of anxiety. Similar as for anxiety, the peaks corresponding with the individual markers are dominated by the \textit{health} marker, followed by \textit{suffering} and \textit{fear}.

Recent evidence has suggested that people who are kept under lock-down experience significant levels of anxiety, anger, confusion, depression and stress~\citep{salari2020prevalence}. In this context, the variations observed in the markers associated to each of the selected mental disorders as the COVID-19 lock-down situation evolved, show that social media reflects the events in the real world, thus helping to answer RQ1. This is further confirmed by the matching between the noticeable changes in the prevalence of markers and relevant COVID events in Argentina. 
These observations are in accordance with those of~\citet{barzilay2020resilience}, who discovered through surveys and questionnaires symptoms of anxiety, depression and stress in the population.  Similarly,~\citet{adams1984mount,nolen1991prospective,jeong2016mental} determined through surveys and interviews changes in stress and depression markers through the crisis period. In addition to the direct COVID-19 situation, the high levels of anxiety and stress could be also related to the economic consequences of lock-down~\citep{mucci2016correlation}. 
Finally, the evolution of the manifestations of the mental health disorders also showed a correspondence with the psychological states of a crisis~\citep{drabek2012human,dewolfe2000training}, by which it is expected to first have an \texttt{anxious} phase, followed by \texttt{stress} and \texttt{depression}.

\subsection{Emotions}

\begin{figure}
\includegraphics[width=0.98\columnwidth]{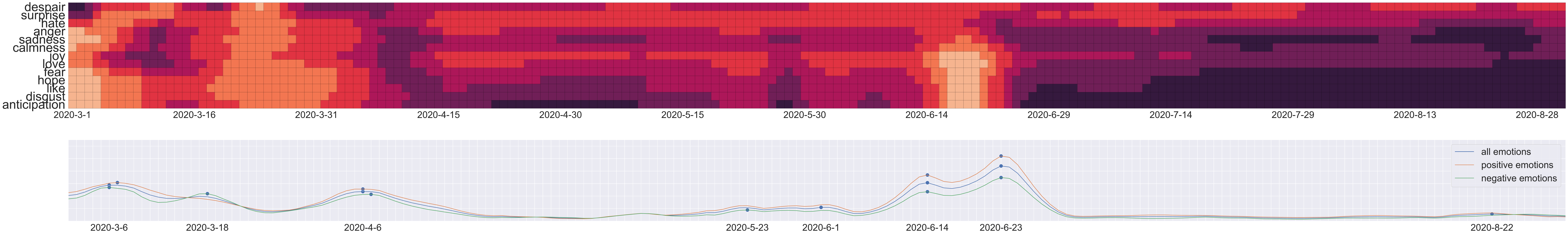}
\caption{Prevalence of the SentiSense emotions over the span March-August}\label{fig:emotions}
\end{figure}

Sentiment analysis has been shown as an effective tool to detect social media content that can contribute to situational awareness, as it can help to understand the dynamics of individuals~\citep{gruebner2018spatio}. For example, how individuals are coping with the causes and effects of the crisis, which are their main concerns, or the emotional burden of the crisis. Nonetheless, some studies~\citep{WHITTLE201260,mort2005psychosocial} have argued that it is not enough to solely focus on mental health issues or a global sentiment score, as they might miss the complexity of the full range of emotional responses to not only the direct effects of crises, but also to the social, economic and environmental changes caused by them. 
Hence, assessing multiple emotion dimensions might provide insights regarding how a crisis or disaster is experienced by the affected individuals~\citep{10.1371/journal.pone.0181233}. In this sense, we consider a subset of the SentiSense \texttt{emotions}, which are shown in Figure~\ref{fig:emotions} for the span March-August, and the detected peaks for both the \textit{positive} (i.e. surprise, calmness, joy, love, hope, like and anticipation) and the \textit{negative} (i.e. despair, hate, anger, sadness, fear and disgust) emotions\footnote{The peak Figures for each of the individual emotions can be found in the companion repository}. The darker the color the higher the prevalence of the emotion for such day. Unlike for the mental health markers, July and August did not show a magnitude order increment in the prevalence of emotions. 

The \texttt{emotion} evolution shows the appearance of areas with a higher prevalence of emotions in the days following the confirmation of the first case until the announcement of the lock-down, affecting both \textit{negative} and \textit{positive} emotions. The \textit{positive} emotions could be related with the mentioned confidence in the decisions made by the government, and at the same time the feeling of being taken care.  At the same time, there was also a small change in \textit{anticipation}, which could be related to a sense of uncertainty regarding what to expect during lock-down. The prevalence of \texttt{emotions} is similar to those of the detected Empath categories, for example in both cases \textit{anger} and \textit{sadness} showed a high prevalence mid-March, around the time of the first official prevention measure and lock-down announcements. 

In early-June the lock-down was extended for its sixth time until the end of June, with new allowed activities for the big cities, including running, individual outdoor sports and shopping. Unlike the analyzed mental health traits, most emotions showed a low prevalence period around mid-June, which could be related to a sense of optimism for the newly authorized activities. Then, late-June contagions increased, leading to the following extension of the lock-down, and the start of the most emotional period. 
When observing the peaks for the \textit{negative} emotions, they are dominated by changes in fear, disgust, and (at a lesser extent) hate, which achieved their biggest changes in early-March (the first COVID related announcements including the first lock-down), and mid- to late- June (when reaching the 100 days of lock-down), which match the changes observed for the mental health traits. Then, \textit{disgust} and \textit{fear} changed in the same magnitude in June. Changes in the remaining emotions were less prominent. \textit{Despair} was the emotion with the smallest changes. Small peaks were observed for most emotions around the dates of the lock-down extension announcements. By July, none of the negative emotions showed prominent peaks, indicating a sustained level of emotions, with no sudden changes. Table ~\ref{tab:peaks-events-emotions} complements Table\ref{tab:peaks-events} matching the discovered emotion peaks with events.

\begin{table}

\caption{Events associated to peaks observed for the emotions analysis\label{tab:peaks-events-emotions}}
\scalebox{0.90}{

\begin{tabular}{p{0.1\columnwidth}p{0.9\columnwidth}}
\toprule 
{\footnotesize{}Peaks} & {\footnotesize{}Associated events}\tabularnewline
\midrule
\midrule 
{\footnotesize{}March 18th} & \textbf{\footnotesize{}March 13th.}{\footnotesize{} The second death
is confirmed.}{\footnotesize\par}

\textbf{\footnotesize{}March 14th.}{\footnotesize{} Schools are to
be closed. Borders are closed to non-residents coming from affected
countries. }{\footnotesize\par}

\textbf{\footnotesize{}March 17th.}{\footnotesize{} Citizens are asked
not to do tourism. }\tabularnewline
\midrule 
{\footnotesize{}April 6th} & \textbf{\footnotesize{}March 31st.}{\footnotesize{} Borders are still
close for non-residents. Lock-down extension for two weeks. }{\footnotesize\par}

\textbf{\footnotesize{}April 2nd.}{\footnotesize{} Protests asking
for the executive branch of the government to reduce their salaries. }{\footnotesize\par}

\textbf{\footnotesize{}April 3rd.}{\footnotesize{} University of Buenos
Aires presents a plan for returning to activities on June 1st. The
judicial branch of the government reduces their salaries.}{\footnotesize\par}

\textbf{\footnotesize{}April 4th.}{\footnotesize{} Mercosur approves
an emergency fund and 45.000 diagnostic tests will arrive.}{\footnotesize\par}

\textbf{\footnotesize{}April 5th.}{\footnotesize{} A TV special ("Unidos
por Argentina", "United for Argentina")
is held to collect donations through the Red Cross. More than 1500
confirmed cases.}{\footnotesize\par}

\textbf{\footnotesize{}April 6th.}{\footnotesize{} Controversy over
purchases made by the government with surcharges.}\tabularnewline
\midrule 
{\footnotesize{}June 14th} & \textbf{\footnotesize{}June 8th.}{\footnotesize{} Changes in the authorized
activities in CABA. The government presents a plan for expropriating
a cereal exporter. }{\footnotesize\par}

\textbf{\footnotesize{}June 14th.}{\footnotesize{} New restrictions
in Buenos Aires. }\tabularnewline
\midrule 
{\footnotesize{}August 22nd}  & \textbf{\footnotesize{}August 17th.}{\footnotesize{} Protest in all
the country against the government. }{\footnotesize\par}

\textbf{\footnotesize{}August 19th.}{\footnotesize{} New protest against
changes in the judicial branch of governement. }{\footnotesize\par}

\textbf{\footnotesize{}August 20th.}{\footnotesize{} 4\% GDP deficit. }{\footnotesize\par}
\bottomrule
\end{tabular}
}

\end{table}

As regards the \textit{positive} emotions, peaks show similar tendencies than those for the \textit{negative} ones. Both early-March and late-June showed the highest peak density and prominence.
Unlike for the negative emotions, in late-July and mid-August there appeared there appeared a few \textit{surprise} and \textit{calmness} small peaks. These peaks match different political and economic events. For example, there is a peak around the celebration of Argentina Independence Day (July 9th) and the day the president presented the plan for recovering the economic situation after the pandemic (July 11th). Then, the \textit{surprise} peaks in August match the new extensions of the lock-down and the authorization of new activities in the capital city.

These observations are in agreement with those of~\citet{gruebner2018spatio} who found differences in the level of discomfort (represented by six negative emotions: anger, confusion, disgust, fear, sadness, and shame) of social media users before, during and after a disaster. Particularly, they observed that \textit{negative} emotions tended to increase during and after the crisis, in contrast with the emotions during the warning or planning phase before the crisis. 
Similarly, \citet{10.1371/journal.pone.0181233} observed a prevalence of \textit{fear} and \textit{surprise} after the disaster (in this case, this could mean the days after the announcement of the first lock-down), and an excess of \textit{sadness}. Additionally, the authors determined that the emotions showing the highest prevalence (they referred to them as emotions showing an excess of risk) were \textit{anger}, \textit{fear}, \textit{sadness}, \textit{disgust} and \textit{surprise}, which match the emotions dominating the peaks in our observations. Moreover, the evolution of emotions shows a behavior similar to the observed by~\citet{aiello2020epidemic} for the \textit{fear} social epidemic in the two-month period between the detection of the first case and the declaration of lock-down, in which there is a prevalence of \textit{fear} and \textit{anger}, followed by \textit{sadness} and some positive emotions. The observation of the changes in the prevalence of emotions and the evidence of their interrelations in agreement with other works in the literature allow to answer RQ2, reinforcing the role of social media as a means for monitoring emotions and, consequently derived negative mental health outcomes.

The evolution of the emotions could be also associated to the psychological states of crisis~\citep{drabek2012human,dewolfe2000training} by which it is expected to have a sense of \textit{anticipation} in the days leading to the prime event of the crisis (in this case, it could include the days inbetween the confirmation of the first case and the announcement of the lock-down), followed by a period of \textit{negative} emotions associated with states of anxiety. Then, there can be brief periods of optimism, which are again followed by a prevalence of negative emotions and stress traits.

\subsection{Crisis}

\begin{figure}
\centering
  \subfloat[Preparedness\label{fig:crisis-preparedness}]{   
     \hspace{-0.2cm}\includegraphics[width=0.98\columnwidth]{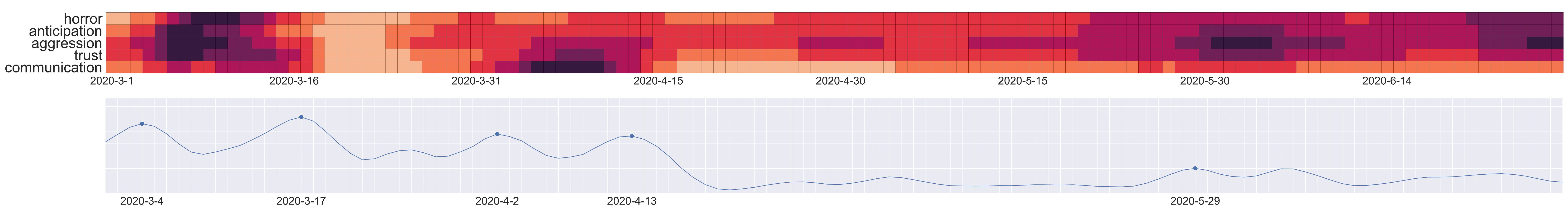}}\\
     
   \subfloat[Response\label{fig:crisis-response}]{
    \hspace{-0.2cm}\includegraphics[width=0.98\columnwidth]{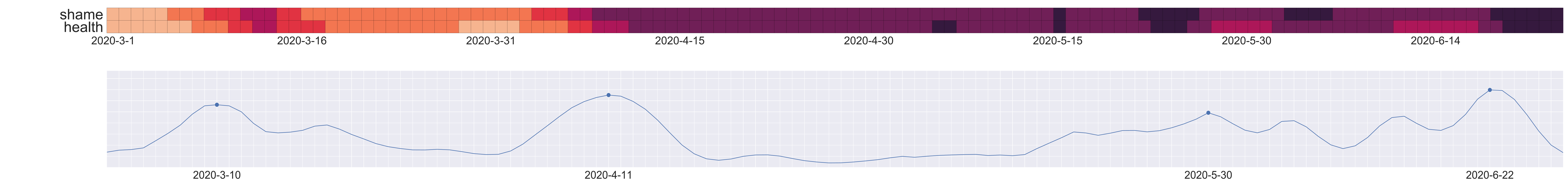}}\\
     
    \subfloat[Recovery\label{fig:crisis-recovery}]{\hspace{0.1cm}\includegraphics[width=0.98\columnwidth]{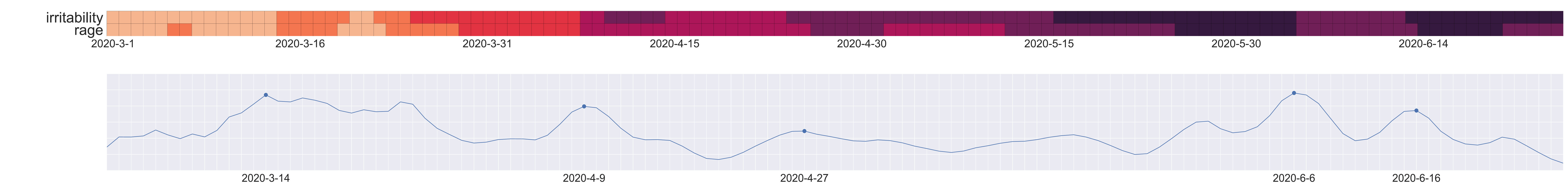}  }\\
   
   \subfloat[Common to \textit{preparedness}, \textit{response} and \textit{recovery}\label{fig:crisis-common}]{\includegraphics[width=0.98\columnwidth]{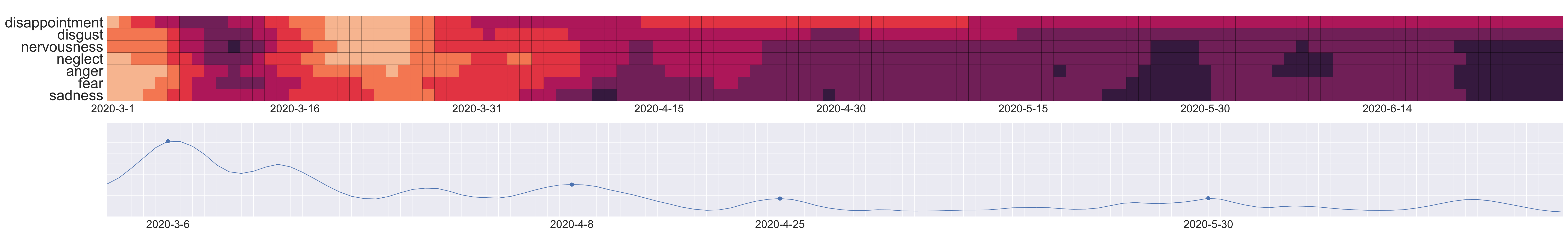}}
   
 \caption{Prevalence of the Empath categories across crisis stages}\label{fig:crisis-stages}
\end{figure}

Based on the described operationalization of the traditional crisis stages~\citet{neal1997reconsidering,drabek2012human}, Figure~\ref{fig:crisis-stages} presents the temporal distribution of references to each of the particular markers associated to the classical disaster stages~\citep{neal1997reconsidering,dewolfe2000training}: \textit{preparedness}, \textit{response}, \textit{recovery}, and the shared markers across stages, for the span March-June, along with the detected peaks. As for the mental health analysis, during July and August we observed different orders of magnitude across the markers, which hinders the observation of changes in the previous months. The superposition of the categories shared by \textit{preparedness}, \textit{response}, \textit{recovery} show their coincidences and their interweaved nature.
The \textit{mitigation} phase is not included in the analysis as it is related to activities after the recovery of the current event, and in preparation of future similar events. Due to the nature of the pandemic, the current state of the health situation in Argentina, and a preliminary analysis of the categories associated to this phase, it does not seem that Argentinian society has yet arrived to this stage. Table~\ref{tab:peaks-events-stages} complements Table~ref{tab:peaks-events} and Table~ref{tab:peaks-events-emotions} matching the discovered stages peaks with events.

\begin{table}[t]

\caption{Events associated to peaks observed for the crises stage analysis\label{tab:peaks-events-stages}}
\scalebox{0.90}{

\begin{tabular}{p{0.1\columnwidth}p{0.9\columnwidth}}
\toprule 
{\footnotesize{}Peaks} & {\footnotesize{}Associated events}\tabularnewline
\midrule

{\footnotesize{}April 13th} & \textbf{\footnotesize{}April 6th.}{\footnotesize{} The president justifies
the ministries that allegedly paid overprices. Mandatory use of masks.
Pope Francis declares that in prisons COVID can cause a "calamity". }{\footnotesize\par}

\textbf{\footnotesize{}April 8th. }{\footnotesize{}More than 800 prisoners
are released for risk of COVID contagion. }{\footnotesize\par}

\textbf{\footnotesize{}April 10th.}{\footnotesize{} Announcement of
the new lock-down extension. Announcement of financial assistance
for tourism related companies and sport clubs. }{\footnotesize\par}

\textbf{\footnotesize{}April 12th.}{\footnotesize{} The president
replies to the Chilean government that number shown that the applied
policies are working. The number of domestic violence reports increases.}\tabularnewline
\midrule 
{\footnotesize{}April 27th} & \textbf{\footnotesize{}April 20th. }{\footnotesize{}The government
announces that 3 millions extra people has filled for food assistance. }{\footnotesize\par}

\textbf{\footnotesize{}April 21st.}{\footnotesize{} Plans for a new
lock-down extension. Crisis and inflation warnings. }{\footnotesize\par}

\textbf{\footnotesize{}April 22nd.}{\footnotesize{} The health minister
affirms that he did not contemplated the possibility of allowing entertainment
activities. }{\footnotesize\par}

\textbf{\footnotesize{}April 25th.}{\footnotesize{} Announcement of
the new lock-down extension. People is now allow to go outside one
hour a day for a walk.}{\footnotesize\par}
\midrule 

{\footnotesize{}April 27th} & \textbf{\footnotesize{}April 20th. }{\footnotesize{}The government
announces that 3 millions extra people has filled for food assistance. }{\footnotesize\par}

\textbf{\footnotesize{}April 21st.}{\footnotesize{} Plans for a new
lock-down extension. Crisis and inflation warnings. }{\footnotesize\par}

\textbf{\footnotesize{}April 22nd.}{\footnotesize{} The health minister
affirms that he did not contemplated the possibility of allowing entertainment
activities. }{\footnotesize\par}

\textbf{\footnotesize{}April 25th.}{\footnotesize{} Announcement of
the new lock-down extension. People is now allow to go outside one
hour a day for a walk. }{\footnotesize\par}
\midrule 
{\footnotesize{}June 6th} & \textbf{\footnotesize{}May 30th.}{\footnotesize{} Debt negotiations
to avoid default. }{\footnotesize\par}

\textbf{\footnotesize{}May 31th}{\footnotesize{}. Government agencies
are being investigated for alleged overpricing. }{\footnotesize\par}

\textbf{\footnotesize{}June 1st. }{\footnotesize{}Over 21 million
people receive financial help from the government. }{\footnotesize\par}

\textbf{\footnotesize{}June 4th.}{\footnotesize{} The president shares
a video announcing the new lock-down extension. }{\footnotesize\par}

\textbf{\footnotesize{}June 5th.}{\footnotesize{} Debt negotiations
are continued to avoid default.}{\footnotesize\par}

\bottomrule
\end{tabular}
}

\end{table}

As regards the \texttt{preparedness} phase (Figure~\ref{fig:crisis-preparedness}, \ref{fig:crisis-common}), as for \textit{anxiety}, the distribution shows the apparition of areas of high prevalence of \textit{anticipation}, \textit{trust} and \textit{nervousness} in the days following the confirmation of the first COVID case and previous to the first official announcements. The peaks for this phase match the areas with high prevalence in March. \textit{Aggression} could be related to anger, as both serve as a self-defensive mechanism for coping with the situation. Anger could also be explained in terms of the reactions to the political events previously described for \texttt{anxiety}. 

When individually analyzing the categories, \textit{communication} presents the most prominent peaks from March until mid-April, matching the time span in which the government made an effort to promote safety measures for preventing contagions, and the announcements of the first extension of lock-down, and the first locally produced COVID tests. 
During this time, the government promoted the idea that Argentina was leading the fight against the COVID and that the efforts were worldwide renown, which could have spiked the sense of trust.  
In June, when the contagions increased again after the first lock-down flexibilization, another \textit{communication} peak appeared, which could be related with the need of reminding individuals of both the prevention measures\footnote{Official government spot: \url{https://twitter.com/msalnacion/status/1272569410407063554}} and the consequences of the disease\footnote{For example, the government created official short videos to discourage people of maintaining social reunions: \url{ https://twitter.com/SantiCafiero/status/1287502460509130752}}.

When observing \texttt{response} (Figure~\ref{fig:crisis-response}, Figure~\ref{fig:crisis-common}), a peak appeared after the confirmation of both the first contagions and the first death. Nonetheless, markers started to show their high prevalence around mid-April (matching the second peak) when lock-down was once again extended, contagions started to gradually grow, and individuals started to be aware of the possibilities of contagion and the consequences of the disease. 
As previously mentioned, the prevalence of \textit{health} could not only be related to the diffusion of prevention guidelines or the description of physic symptoms or other health complications, but also to the phenomenon of health anxiety. 

According to~\citet{Richardson2005ThePO}, during a crisis, people might suffer “survivor’s guilt”, which might manifest by \textit{shame} for still having a job, for being able to work (the lock-down in Argentina prohibited all activities but a small subset of exceptions), which might explain the high relevance of \textit{shame}. Then, \textit{shame} or even self-guilt can be followed by a period in which there is a need to blame someone. In this case, around mid- and late-June, when several activities were allowed in CABA, the culprit of the new contagions were the runners, which were the target of a blaming campaign in social media. This situation matches a detected peak. Then, the remaining dark areas match the spans over which the president make announcements regarding the health situation in Argentina, and extended the lock-down. 

\begin{figure}
\includegraphics[width=0.98\columnwidth]{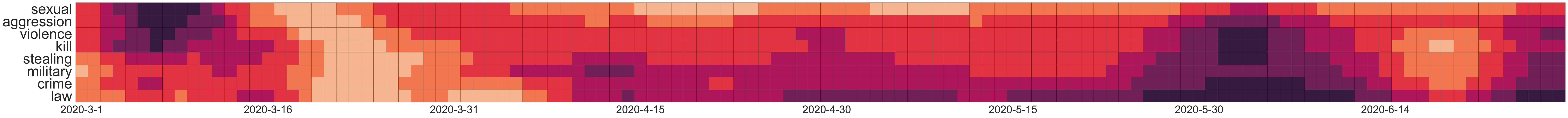}
\caption{Prevalence of the Empath categories associated to \textit{crime} over the span March-June}\label{fig:crime}
\end{figure}

Several theories have explored the relationship between crisis and the prevalence of crime~\citep{Prelog2016ModelingTR}. On one hand, there are theories (such as Therapeutic Community) that support the claim that both violent and property crime either remain unaffected or decrease after crises, while domestic violence could rise. On the other hand, other theories (such as Social Disorganization) claim that crisis have the potential to create or aggravate the existing social disorganization by disrupting the social cohesion and the collective efficacy rendering the community unable to self-monitor and sanction antisocial behavior. In turn, this implies that the rate of criminality tends to rise due to the lack of controls. There is a third group of theories (such as Routine Activity) that suggest that changes in crime rates are the result of changes in the socio-structural organization of everyday activities, reflecting the convergence of three necessary elements: the availability of crime targets, the absence of guardians (such as the police) and the presence of motivated offenders. An analysis of the Empath categories related to crime (shown in Figure~\ref{fig:crime}) revealed that crime did not seem to be a preoccupation during the early stages of lock-down, which is in line with the government declarations in late-March and mid-April claiming that stealing was reduced due to the decrease in movements in the cities. Nonetheless, during late-April and May the prevalence of the categories increased (along with the category related to government mentions), which could be associated the release of prisoners for fear of contagion, and the rise in crimes against the property by which individuals started the usurpation of private land\footnote{This rise in land usurpation led to the political problem that until October 2020 has not been solved.}. Later in August, the government admitted a rise in criminality.

In social terms, the \texttt{recovery} stage (Figure~\ref{fig:crisis-recovery}, Figure~\ref{fig:crisis-common}) covers the attempts to return to a “normal life” mixed with fear, anxiety, depression, rage, irritation with changes in the daily life, and the appearance of conflicts at the community level, among other characteristics~\citep{drabek2012human}. In this sense, for this stage, there are two non-overlapping markers, \textit{irritability} and \textit{rage}, which showed a high and continuous prevalence starting in mid-May. By late-May, Argentina had already endured 5 lock-down extensions and the number of contagions was still on the rise. In parallel, the press conferences led by the government included comparisons to other countries to show how well the health situation was been administered (e.g. Sweden, Chile, Spain, and Brazil), which were debunked. 
The fact that the markers associated to \texttt{response} and \texttt{recovery} show a high prevalence around the same time spans shows the interweaved nature of the phases and that recovery actions were followed while society was still enduring the impact of the pandemic. For example, while the number of contagions was still rising, the government tried to better equip the health system by providing more budget and equipment to the most affected regions.

The high prevalence of the markers associated to the three stages during July and August could be perceived as a return to the early stages of the pandemic as, at that time, the reporting of contagions and deaths as well as the positivity rate started to rise. At the same time, the economic situation continued to aggravate as several sectors were unable to resume working. This situation is then expressed in social media by the observed exacerbated prevalence of markers associated to mental health and emotions, as previously showed.

According to the observed changes in social media activity, the identified mental states and emotions, and the stationary prevalence of the derived markers, it can be stated that Argentina has traversed the three classical stages of crises, thus providing evidence for answering RQ3. Based on the analysis, the \texttt{preparedness} spanned between early-March and mid-April, covering the time when society was still trying to make sense of the situation, and the first prevention measures were adopted after the first cases, deaths and the first lock-down periods.
Then, the \texttt{response} stage started mid-April, when the lock-down was well established, the development of national tests and the announcement of the first economic and political measures were announced, for example, the creation of a universal salary for helping those that were not able to work under the restrictions. Finally, while continuing in the \texttt{response} stage, a brief \texttt{recovery} stage started in late-May and early-June, with the first lift of restrictions in the big cities, signaling the first brief steps towards a new normality. This situation was then reverted in July and \texttt{response} took prevalence again as the number of daily contagions rose and more restrictive lock-downs were applied. 

To further reinforce both the differences and relations between the three stages, Table~\ref{tab:category-prev} reports the maximum percentage difference between the prevalence of the marker in each of the stages compared to the median prevalence of the marker across the span March-June. In each row, the maximum value is highlighted. As the Table shows, in most cases the highest percentage difference of a marker is observed for its corresponding stage. The two exceptions are \textit{aggression} and \textit{health}, both showing a higher prevalence in the \texttt{recovery} stage. For those markers that are shared across stages, the average prevalence differences across phases are lower, i.e. the category usage is more evenly distributed across phases than for those categories associated to a unique phase.

\begin{table}
\caption{Prevalence of the Empath categories over the crisis stages between March-June\label{tab:category-prev}}
\centering
\begin{tabular}{ccccc}
\toprule 
 &  & \textit{preparedness} & \textit{response} & \textit{recovery}\tabularnewline
\midrule 
\multirow{5}{0.32\textwidth}{Associated to \textit{preparedness}} & horror & \textbf{52.45} & 9.33 & 17.83\tabularnewline
 & aggression & 13.53 & 7.27 & \textbf{15.50}\tabularnewline
 & anticipation & \textbf{41.02} & 7.95 & 22.62\tabularnewline
 & communication & \textbf{27.46} & 1.72 & 4.06\tabularnewline
 & trust & \textbf{20.05} & 2.81 & 6.99\tabularnewline
\midrule 
\multirow{2}{0.32\textwidth}{Associated to \textit{response}} & shame & 13.78 & \textbf{23.74} & 16.41\tabularnewline
 & health & 12.59 & 28.93 & \textbf{39.58}\tabularnewline
\midrule 
\multirow{2}{0.32\textwidth}{Associated to \textit{recovery}} & irritability & 11.64 & 23.75 & \textbf{24.20}\tabularnewline
 & rage & 3.22 & 16.03 & \textbf{27.10}\tabularnewline
\midrule 
\multirow{5}{0.32\textwidth}{Associated to \textit{preparedness}, \textit{response} and \textit{recovery}} & fear & \textbf{28.33} & 12.47 & 14.17\tabularnewline
 & sadness & 14.06 & \textbf{17.80} & 11.44\tabularnewline
 & nervousness & \textbf{25.08} & 16.46 & 18.60\tabularnewline
 & neglect & 9.96 & 16.62 & \textbf{20.52}\tabularnewline
 & disappointment & \textbf{36.27} & 5.97 & 6.57\tabularnewline
\midrule 
\multirow{2}{0.32\textwidth}{Associated to \textit{response} and \textit{recovery}} & anger & 4.00 & 16.00 & \textbf{21.70}\tabularnewline
 & disgust & 6.16 & 12.39 & \textbf{18.90}\tabularnewline
\bottomrule 
\end{tabular}

\end{table}

\section{Conclusions}\label{sec:conclusions}

The COVID-19 pandemic has profoundly affected all aspects of society, not limited to the physical health, but also to mental health, economics (e.g. affecting employment conditions, financial insecurity and poverty) and even twisting political decisions. 
At the time of writing this paper, Argentina has just endured 200 days of lock-down, while being at the top-10 countries with most contagions. Even though some restrictions have been lift, there still holds the restriction of freely moving around the country, and students have not yet returned to schools or universities. 
In this context, measuring the effects of the pandemic on individuals and societal dynamics is vital to understand the policies used to manage the pandemic, which should aim at achieving a right balance between the disease control and the mitigation of the negative socio-economic effects~\citep{holmes2020multidisciplinary}. Moreover, government agencies should provide accurate information on the state of the pandemic, refute rumours in a timely manner and reduce the impact of misinformation, which should result in a sense of public security and trust~\citep{salari2020prevalence}.

Most studies on the effect of crises and disasters have solely relied on post-disaster data, which were obtained through surveys or questionnaires~\citep{gruebner2018spatio} and conducted in a small-to-medium scale, which may cause important information to be missed. We believe that social media can complement these studies by enabling opportunities to track individual behavior and perceptions over longer periods of time, from pre- to post-crisis, and at a large scale. Moreover, analyzing social media can also help to monitor the spread of the disease, the awareness about the disease and symptoms, and the responses to health and government recommendations and policies, thus helping in the design of effective communication interventions~\citep{van2020using,depoux2020pandemic} to provide reassurance and practical advice. 

From a theoretical point of view, our analyses provide a contextualization of the traditional stages of a crisis, showing their corresponding manifestations in social media, which in turn verifies the processes and patterns previously observed through surveys. In addition, it helps to discover mental health markers, which are necessary for the design of health prevention policies~\citep{holmes2020multidisciplinary}. The analysis approach presented in this study can, in principle, be applied to any large-scale data stream and have practical implications on the monitoring of individuals’ behaviors, perceptions, mental health and emotions, which are useful indicators to understand mechanisms and propose interventions.  Based on the independence of the defined lexicons with respect to the nature of a crisis (i.e. the lexicons do not reflect particularities of the crisis at hand, but rather they characterize mental health markers and the generic stages of crises), the analysis, process and techniques can be applied to future crises and disasters, provided that social media data can be collected.

There are also interesting aspects to consider in future works. First, we characterized the COVID-19 pandemic in the context of Argentina, disregarding its effect over neighbour countries. It should be possible to compare how the pandemic manifested in other countries and whether the different policies adopted by each government, its political orientation and the cultural dimensions, can condition the evolution of emotions and crisis stages. Second, we should continue the analyses to include the current developments, and after life returns to the “new normality”, we could better assess the pandemic long-term effects. Third, even though we focused the analyses on Twitter data, the government has also an official presence on other social media sites, in which the same content is shared, but the individuals consuming it could be different. A challenge here is how to include the different sites in the analyses to account for different perspectives and extend the scope of the analyses. In turn, it would be possible to compare whether users in different sites behave similarly. Fourth, there are ethical concerns regarding the assessment of mental health markers in social media~\citep{GUNTUKU201743}. Even though all collected data is publicly available, there could be risks associated to users’ privacy. If not properly managed, the interventions based on these analyses could lead to discrimination, stigmatization and violence. Hence, a clear policy of data transparency and protection should be defined and enforced. Fifth, given the existence of cases of fake news that were propagated (even by the government) during the early stages of the pandemic, it would be interesting to analyze the prevalence and propagation of misinformation and disinformation, and how they affect the public perceptions and the coping of the crisis. 

\bibliographystyle{plainnat}
\bibliography{references}

\end{document}